\documentclass[12pt]{article}
\usepackage{graphicx}
\usepackage{amsmath,amsthm,amssymb,mathrsfs,mathtools}
\usepackage[OT1]{fontenc} 
\usepackage{etoc}
\usepackage{float}
\usepackage{fancyhdr}

\def\subsectiontitle{}
\def\subsubsectiontitle{}
\usepackage{tikz}
\usetikzlibrary{automata, positioning}

\usepackage{enumitem}
\usepackage[margin=1in]{geometry}
\usepackage[colorlinks,breaklinks=true,pagebackref]{hyperref}
\usepackage{breakcites}

\usepackage{xcolor}
\AtBeginDocument{%
	\hypersetup{%
    linkcolor={red!50!black},%
    citecolor={blue!50!black},%
    urlcolor={blue!80!black}%
}%
}%

\usepackage[nameinlink,noabbrev,sort,capitalise]{cleveref}
\makeatletter
\def\ps@pprintTitle{%
 \let\@oddhead\@empty
 \let\@evenhead\@empty
 \def\@oddfoot{\emph{Very preliminary version}\hfill\emph{This draft: \today}}%
 \let\@evenfoot\@oddfoot}
\makeatother
\usepackage{etoolbox}
\patchcmd{\pprintMaketitle}
  {\fi\hrule}
  {\fi\ifvoid\extrainfobox\else\unvbox\extrainfobox\par\vskip10pt\fi\hrule}
  {}{}

\newsavebox\extrainfobox

\usepackage{inconsolata} 

\newtheorem{prop}{Proposition}
\crefname{prop}{Proposition}{Propositions}

\newtheorem{thm}{Theorem}
\crefname{thm}{Theorem}{Theorems}

\crefname{cor}{Corollary}{Corollaries}

\newtheorem{lem}{Lemma}
\crefname{lem}{Lemma}{Lemmas}

\newtheorem{ass}{Assumption}
\crefname{ass}{Assumption}{Assumptions}

\newtheorem{defi}{Definition}
\crefname{defi}{Definition}{Definitions}

\theoremstyle{remark}

\theoremstyle{definition}

\crefname{eg}{Example}{Examples}

\crefname{problem}{Problem}{Problems}

\usepackage{indentfirst}

\allowdisplaybreaks

\let\oldfootnote\footnote
\renewcommand\footnote[1]{\oldfootnote{\hspace{.4mm}#1}}

\renewcommand{\footnotesize}{\scriptsize}

\makeatletter
\renewenvironment{proof}[1][\proofname] {\par\pushQED{\qed}\normalfont\topsep6\p@\@plus6\p@\relax\trivlist\item[\hskip\labelsep\bfseries#1\@addpunct{.}]\ignorespaces}{\popQED\endtrivlist\@endpefalse}
\makeatother

\let\oldFootnote\footnote
\newcommand\nextToken\relax

\renewcommand\footnote[1]{%
    \oldFootnote{#1}\futurelet\nextToken\isFootnote}

\newcommand\isFootnote{%
    \ifx\footnote\nextToken\textsuperscript{,}\fi}

\usepackage{blkarray}
\usepackage{graphicx,pgfpages,epsfig,multirow,lscape,textcomp}
\usepackage{hyperref}
\usepackage{epic}
\usepackage{xcolor}
\usepackage{setspace}
\usepackage{tikz}
\usetikzlibrary{arrows,arrows.meta,decorations,decorations.pathreplacing,calc,matrix}
\usepackage{natbib}
\begin{document}

\onehalfspacing	

\title{\bf Statistical Discrimination in Ratings-Guided Markets\thanks{Che and Kim are supported by the Ministry of Education of the Republic of Korea and the National Research Foundation of Korea (NRF-2020S1A5A2A03043516).}} 
\author{Yeon-Koo Che\thanks{Che: Columbia University, yc2271@columbia.edu.}\hspace{1.5cm}Kyungmin Kim\thanks{Kim: Emory University, kyungmin.kim@emory.edu.}\hspace{1.5cm}Weijie Zhong\thanks{Zhong: Stanford University, weijie.zhong@stanford.edu.}}
\date{November 2024}
\maketitle

\begin{abstract}  We study statistical discrimination of individuals based on payoff-irrelevant social identities in markets that utilize ratings and recommendations for social learning. 
Even though rating/recommendation algorithms can be designed to be fair and unbiased, ratings-based social learning can still lead to discriminatory outcomes. Our model demonstrates how users' attention choices can result in asymmetric data sampling across social groups, leading to discriminatory inferences and potential discrimination based on group identities.


\end{abstract}
\newpage

\section{Introduction}
Discrimination based on race, gender, ethnicity, and other social identities remains a persistent societal challenge.   This age-old problem has not vanished in the digital era; it continues to appear in modern online marketplaces and social media platforms.   Evidence suggests that discrimination is prevalent in popular online platforms such as Airbnb \citep{edelman2017racial,cui2019reducing}, freelancing websites \citep{hannak2017bias}, ride-sharing platforms \citep{ge2016}, and even online communities like math Stack Exchange \citep{bohren2019dynamics}.

At first glance, discrimination in online marketplaces is rather surprising, given their widespread use of user-generated rating systems.\footnote{Platforms for ride-sharing, freelancing, credit, and insurance collect data on users' past performances and experiences, aggregating it into \emph{ratings} to make \emph{recommendations}. Machine-learning algorithms now routinely pre-screen résumés, evaluate loan and insurance applicants, assess freelance workers, suggest promotions or terminations, and predict parolee recidivism.}  Such rating systems  facilitate social learning, which on balance should limit the potential for statistical discrimination. With more accurate information about {\it individuals}, there should be less room for biased inferences based on their {\it group identities.} Indeed, in a world of perfect information, discrimination, except perhaps based on personal preferences, should vanish.\footnote{There is some empirical evidence that reputation/ratings ameliorate discrimination in some contexts \citep[see][]{cui2019reducing,bohren2019dynamics}.}


However, it is not self-evident that more information and more social learning always bring less discrimination.   Crucially, it is unclear whether the social learning mechanisms inherent to these marketplaces operate in a way that reduces discrimination. In essence, social learning involves feedback between two processes: (1) \emph{data sampling (or experience gathering)} and (2) \emph{informing (or recommending) user decisions.} The latter process can be designed to be fair and unbiased, in keeping with the recent emphasis on algorithmic fairness. However, the former process is inherently non-random and potentially biased. Data sampling occurs when transactions take place, driven by the economic interests of the parties involved. Users naturally seek high-value partners with positive ratings, not random or representative ones. Without understanding the nature of selective sampling induced by market forces, one cannot truly understand the fairness of social learning and its implications for discrimination.




This paper develops a model of social learning that incorporates this feedback process to examine its implications for statistical discrimination.  We specifically investigate whether and to what extent social learning through ratings can alleviate or exacerbate discrimination.

Our model features directed search/matching between two sides, buyers and sellers, guided by user-contributed ratings.   Each seller is indexed by her social group identity $\ell=1,2$ and her productivity type, $H$ (``high'') or $L$ (``low'').  While a seller's group identity remains constant, their productivity can change over time according to a continuous-time Markov process. Importantly, group identity is assumed to be independent of a seller's productivity. Buyers aim to match and trade with sellers, and the surplus generated from trade is higher when the seller is of type $H$ than type $L$.
  
  The search-matching process is frictional and guided by imperfect information about sellers' types, called {\it ratings}.  The ratings are binary, $G$ (``good'') or $B$ (``bad''), and are updated after each trade.  Although it is impossible for the ratings to perfectly reveal sellers' types due to their ever-changing nature (except in the limit), we capture the effectiveness of social learning---the degree to which ratings reflect actual types---using a parameter $\alpha\in (0,1)$. This parameter represents the probability that an incorrect rating is corrected after a transaction. Buyers strategically direct their search attention based on both ratings and group identities. The probability of a seller matching with a buyer depends on the number of buyers focusing on the sellers with a given rating and group combination. We analyze the steady-state behavior of this system.



 The payoff irrelevance of group identity means that there always exists a ``non-discriminatory'' equilibrium where sellers from both groups are treated identically. In this equilibrium, sellers receive the same level of attention from buyers regardless of their group identity, with $G$-rated sellers enjoying a higher match rate than $B$-rated sellers due to the positive signal associated with a good rating.  This equal treatment ensures unbiased sampling across groups, leading to identical belief updates. Consequently, non-discrimination can be maintained and persist in a steady state.

 However, a non-discriminatory equilibrium need not be the only equilibrium, or a stable one.\footnote{As will be seen, stability is defined in the usual manner, by the robustness of an equilibrium to small perturbations.}  A discriminatory equilibrium can also arise due to a potential positive feedback loop between data sampling and belief updating: sellers with favorable ratings attract more matches, leading to more frequent sampling and updating. To illustrate, suppose buyers favor $G$-rated sellers in social group 1 ($G1$) over $G$-rated sellers in social group 2 ($G2$). Then, $G1$ sellers will be matched more frequently, leading to more accurate ratings and a stronger filtering out of low-type sellers. This, in turn, reinforces the favorable perception of $G1$ sellers over $G2$ sellers, justifying the initial bias. This feedback loop leads to a discriminatory equilibrium where buyers prefer $G1$ sellers over $G2$ sellers. Interestingly, this feedback loop does not apply to $B$-rated sellers ($B1$ and $B2$). If $B1$ sellers receive more attention and match more frequently, their ratings become more accurate, but this leads to a less favorable perception, correcting the initial bias. Thus, in the steady state, $B1$ and $B2$ sellers are necessarily treated equally.\footnote{This differential effect across different ratings resembles a ``glass ceiling'' --- larger wage and participation gaps at the top end of the labor market, which is a well-documented phenomenon in labor markets based on both gender and race. See, e.g.,  \cite{mcdowell1999cracks,bertrand2004emily,lehmann2011job,bartovs2016attention,green2021measuring,bertrand2019breaking,field2020table}.}

In summary, this discriminatory equilibrium favors sellers from one group over another, even though group identity has no impact on payoffs and there is no inherent bias in the rating algorithm or belief updating process. While a discriminatory equilibrium might not always exist, when it does, the non-discriminatory steady state is often  unstable: even a slight shift in buyer attention towards one group can disrupt the non-discriminatory equilibrium and lead to a discriminatory one.  By contrast, a stable discriminatory equilibrium exists in such a case.

Our analysis demonstrates that a discriminatory equilibrium can emerge when matching and social learning frictions are at an intermediate level.  
This suggests that advancements in online marketplaces, reducing market frictions, may have   a non-monotonic effect on discrimination.  Initially, high frictions might support only a non-discriminatory equilibrium. However, as matching and social learning improve, these frictions diminish, potentially allowing a discriminatory equilibrium to emerge.

There is a  long line of research on how discrimination may arise from rational statistical inferences on groups' characteristics.\footnote{Another line of research considers a direct taste-based discrimination.  See \cite{becker1957}.}   Unlike seminal works focusing on exogenous group differences (\cite{phelps1972} and \cite{arrow1973theory}),  modern theories of statistical discrimination focus on feedback loop between groups' endogenous human capital choices and market beliefs about these choices (\cite{coate1993antidiscrimination}, \cite{mailath2000endogenous}, \cite{norman/moro:2003}, \cite{norman2003statistical}; see a survey by \cite{fang2011theories}): if employers view a certain group as less skilled and thus become more selective against them for assigning higher-paying positions, the affected group will indeed lose incentives for acquiring skills, thus fulfilling the employers' adverse beliefs on that group.  More recent papers consider endogenous information channel:  \cite{gu2020searchmodelstatisticaldiscrimination} attribute statistical discrimination to endogenous occupational choice. \cite{echenique2022rationally} study the strategically inattentive behavior of employers stemming from coordination failures in worker investment.\footnote{\cite{cornell1996culture} explain discrimination and its inter-generational persistence from  group-specific evaluational familiarity.  \cite{bohren2019dynamics} and \cite{monachou2019discrimination} focus on the ``mis-specified'' prior beliefs as a source of discrimination.}  

The current paper explores a distinct mechanism for discrimination. First, discrimination emerges as a rational response to social learning on sellers and does not require endogenous human capital acquisition.  Existing theories focus on how anticipated {\it future} discrimination affects skill investment, validating discrimination in equilibrium. In contrast, our model highlights how {\it past} discrimination influences buyers' inferences about sellers' good ratings, perpetuating discrimination.  Second, unlike existing theories where discriminatory equilibria rely on multiple non-discriminatory equilibria,   discriminatory equilibria may exist even when there is a unique non-discriminatory equilibrium. Finally, non-discriminatory equilibria are stable in existing theories but are often unstable in ours when a discriminatory equilibrium exists, making discrimination a more robust prediction in our model.\footnote{While less directly related, \cite{bardhi2020early} adopt a similar stability analysis to study the long run impact of minor early-career discrimination through dynamic learning, showing that non-discriminatory outcomes are stable under success-tracking but become unstable under failure-tracking.}

Our paper also connects with the growing literature in computer science on ethical algorithms \citep[see][among others]{dwork2012fairness,corbett2017algorithmic,kearns2018preventing,kleinberg2018discrimination}. This literature investigates methods to ensure that decision/recommendation algorithms adhere to various fairness standards. Our work qualifies the effectiveness of this approach by highlighting that algorithmic fairness alone might not be sufficient to achieve true fairness. In our model, even if the rating algorithm treats both groups equally, a discriminatory equilibrium can still emerge if agents (buyers) \emph{interpret} ratings in a way that leads to discriminatory sampling. The debate on algorithmic fairness must consider this aspect of social learning. Either the scope for interpretation needs to be completely eliminated to guarantee fair outcomes, an approach that appears to be in line with the prescription of \cite{kleinberg2018discrimination}, or the rating system itself should be designed to counteract potential biases in user interpretation.  

\section{Model}

We consider a decentralized market in which buyers (or firms) search for sellers (or workers) of unknown types. 

\paragraph{Players.} There is a unit mass of sellers in the market. The sellers are indexed by two characteristics, {\it type} and {\it group}. The type of a seller represents any payoff-relevant information, such as the productivity or quality of the seller. At a given moment, a seller is either of  high type ($H$) or low type ($L$). Each seller's type, however, changes according to a continuous-time Markov process. Specifically, each type turns into the other type at rate $\delta>0$.\footnote{If $\delta$ tends to $\infty$ then past records (so ratings) become irrelevant for the market outcome. If $\delta$ tends to $0$ then either the market completely breaks down (see \cref{prop:trade_vs_no_trade}) or ratings fully reveal sellers' (perfectly persistent) types.} 
The group of a seller describes her payoff-irrelevant identity, such as her gender, ethnic or racial identity. Each seller belongs to either group $1$ or $2$, with $\ell=1,2$ as the generic index. Unlike her type, a seller's group does not change over time. For simplicity, we assume that both groups have the same total size (i.e., each group has mass $1/2$), except in \cref{ssec:asymmetry}. 

On the other side of the market, there is mass $Q(>0)$ of buyers. They search for sellers based on public information about sellers. Specifically, they condition their search on sellers' two observable characteristics, rating $j=G,B$ and group identity $\ell=1,2$. 

The sellers who share the same observable characteristics, $(j,\ell)$, and the buyers that search for them constitute a ``submarket,'' which can be indexed by $(j,\ell)$. Sellers are assigned to those submarkets according to their (perfectly persistent) group identity and (evolving) ratings, while buyers choose which submarket to enter.

\paragraph{Matching.} 

We employ the canonical search-and-matching framework to model interactions between buyers and sellers. The matching technology is uniform across all submarkets and exhibits constant returns to scale; in other words, players' matching rates depend only on the ratio of buyers to sellers, so-called ``queue length,'' in their submarket. To be specific, let $P_{j}^{\ell}$ denote the measure of sellers and $Q_{j}^{\ell}$ the measure of buyers in submarket $(j,\ell)$. Then, the associated queue length is $\lambda_{j}^{\ell}:=Q_{j}^{\ell}/P_{j}^{\ell}$. For tractability, we focus on the case where buyers' and sellers' matching rates are $\phi(\lambda_{j}^{\ell}):=(\lambda_{j}^{\ell})^{k-1}$ and $\psi(\lambda):=(\lambda_{j}^{\ell})^k$, respectively for $k\in(0,1)$.\footnote{\label{fn:Cobb_Douglas}These matching rates correspond to the Cobb-Douglas production function $f(Q_{j}^{\ell},P_{j}^{\ell})=(Q_{j}^{\ell})^{k}(P_{j}^{\ell})^{1-k}$. For example, $\phi(\lambda_{j}^{\ell})$ is obtained by dividing $f(Q_{j}^{\ell},P_{j}^{\ell})$ by the measure of buyers, $Q_{j}^{\ell}$.}
Observe that $\psi(0)=0$, $\lim_{\lambda\to\infty}\psi(\lambda)=\infty$, $\psi^{\prime}(\lambda)>0$, and $\psi^{\prime\prime}(\lambda)<0$. In addition, $\phi(0)=\infty$, $\phi^{\prime}(\lambda)<0$, and $\phi^{\prime\prime}(\lambda)>0$. 
As becomes clear later, our results rely mostly on these standard properties of the matching functions and so generalize far beyond this parametric model.

\paragraph{Trade.}

Once a buyer and a seller meet, they transact instantaneously and go back to the market. 
The transaction yields surplus  $u_H$ if the seller's type is $H$ and $u_L$ if the seller's type is $L$, where $u_H>u_L\geq 0$. If a buyer transacts with a seller, the buyer pays $p(\geq 0)$ to the seller.   To exclude trivial cases, we assume $u_{H}>p$ (so that there are gains from trade when a seller is of type $H$), but impose no restriction on the relationship between $u_{L}$ and $p$. Note that our assumption on instantaneous transaction makes our model better applicable for one-shot relationships, common in gig/free-lance economies but can be relaxed at the cost of significant technical complexity (but no conceptual difficulty). 



\paragraph{Ratings.}   Market accumulates information about sellers through simple summary indices, called ``ratings.'' There are two possible ratings:  $G$ (as in ``good'') and $B$ (as in ``bad'').  When searching for sellers, buyers observe only their current ratings; no information about sellers' underlying types or past rating history is available. After each transaction, the seller's rating may be updated to reveal her type. Specifically, with probability $\alpha \in (0,1]$,   a $B$-rated seller with  type $H$ earns $G$ rating, and a $G$-rated seller with type $L$ receives $B$ rating. With the remaining probability $1-\alpha$, the seller's rating stays unchanged. A seller with the correct rating keeps the same rating. Note that a correct rating may turn inaccurate due to the changing type.

\paragraph{Belief and Search Decision.} Buyers' beliefs over a seller's type depend on the rating and the (equilibrium) behavior of all players in the system. Let $P_{ij}^\ell$ denote the measure of sellers with type $i$ and $\mu_j^{\ell}\in[0,1]$ denote the proportion of type $H$ sellers in submarket $(j,\ell)$. Bayes' rule implies that whenever $P_{Hj}^{\ell}+P_{Lj}^{\ell}>0$,
\begin{align*}
    \mu_j^{\ell}=\frac{P_{Hj}^{\ell}}{P_{Hj}^{\ell}+P_{Lj}^{\ell}}.
\end{align*}
Given the belief $\mu_j^{\ell}$ and queue length $\lambda_j^{\ell}$, buyers' expected utility from searching in submarket $(j,\ell)$ is
\begin{align*}
    \phi(\lambda_j^{\ell})(\mu_j^{\ell}u_H+(1-\mu_j^{\ell})u_L-p).
\end{align*}
Note that, since transaction is instantaneous and they do not exit the market, buyers simply maximize their flow utility.

 \paragraph{Solution Concept.} We consider a steady state of the economy in terms of the measures of sellers of different types, ratings and group identities. 
The tuple $\{(P_{ij}^{\ell}, Q_{j}^{\ell})\}_{ i=H,L, j=G,B}^{\ell=1,2}$ constitutes an equilibrium if the following conditions hold:

\begin{itemize}
    \item \textbf{Stationarity}: The inflow and outflow to each status $(i,j,\ell)$ are identical, that is, 
    \begin{equation}
            \begin{split}
P_{HG}^{\ell}\delta &=P_{LG}^{\ell}\delta+P_{HB}^{\ell}\psi(\lambda_B^{\ell})\alpha,	\\
P_{LG}^{\ell}(\delta+\psi(\lambda_G^{\ell})\alpha)&=P_{HG}^{\ell}\delta,\\
P_{HB}^{\ell}(\delta+\psi(\lambda_B^{\ell})\alpha)&=P_{LB}^{\ell}\delta,\\
P_{LB}^{\ell}\delta&=P_{HB}^{\ell}\delta+P_{LG}^{\ell}\psi(\lambda_G^{\ell})\alpha,
\end{split}\label{eq:stationary}
    \end{equation}
where $\lambda_j^{\ell}=\frac{Q_j^{\ell}}{P_{Hj}^{\ell}+P_{Lj}^{\ell}}$. For example, the first line of \eqref{eq:stationary} requires the stationarity of the measure of $(H,G,\ell)$-sellers: the left-hand side represents the outflow of $(H,G,\ell)$-sellers becoming $(L,G,\ell)$, while the right-hand side captures the corresponding inflow, which consists of $(L,G,\ell)$ sellers transitioning to $(H,G,\ell)$ and $(H,B,\ell)$-sellers receiving the $G$-rating after transactions. 

    \item \textbf{Optimality}: Buyers actively search only in submarkets with the highest payoff, that is, $Q_{j}^{\ell}>0$ only when 
    \begin{align*}
    \phi(\lambda_j^{\ell})(\mu_j^{\ell}u_H+(1-\mu_j^{\ell})u_L-p)
\end{align*}
is \emph{non-negative} and \emph{maximized} across all submarkets $(j,\ell)$. 

\item \textbf{Market clearing}: For each $\ell=1,2$, $\sum_{ij}P_{ij}^\ell=\frac{1}{2}$. If buyers' expected payoffs are strictly positive, then $\sum_{j\ell}Q_j^\ell=Q$. 

\end{itemize}
We say that an equilibrium is \emph{non-discriminatory} if $\lambda_j^1=\lambda_j^2$ for each $j$, and \emph{discriminatory} otherwise.

\section{Equilibrium Characterization}

\subsection{Steady state distributions and beliefs}\label{subsec:preliminaries}

We first derive the steady state distribution and beliefs for given queue lengths in each submarket. For $\lambda_{G}^{\ell}+\lambda_{B}^{\ell}>0$, \eqref{eq:stationary} and market clearing imply 
\begin{eqnarray*}
P_{HG}^{\ell}&=&\frac{\psi(\lambda_B^{\ell})(\delta+\psi(\lambda_G^{\ell})\alpha)}{4\left( \delta(\psi(\lambda_G^{\ell})+\psi(\lambda_B^{\ell}))+\alpha\psi(\lambda_G^{\ell})\psi(\lambda_B^{\ell}) \right)},~P_{LG}^\ell=\frac{\psi(\lambda_B^{\ell})\delta}{4\left( \delta(\psi(\lambda_G^{\ell})+\psi(\lambda_B^{\ell}))+\alpha\psi(\lambda_G^{\ell})\psi(\lambda_B^{\ell}) \right)},\\
P_{HB}^\ell&=&\frac{\psi(\lambda_G^{\ell})\delta}{4\left( \delta(\psi(\lambda_G^{\ell})+\psi(\lambda_B^{\ell}))+\alpha\psi(\lambda_G^{\ell})\psi(\lambda_B^{\ell}) \right)},~P_{LB}^\ell=\frac{\psi(\lambda_G^{\ell})(\delta+\psi(\lambda_B^{\ell})\alpha)}{4\left( \delta(\psi(\lambda_G^{\ell})+\psi(\lambda_B^{\ell}))+\alpha\psi(\lambda_G^{\ell})\psi(\lambda_B^{\ell}) \right)}.
\end{eqnarray*}
The corresponding beliefs are $\mu_{G}^{\ell}=\mu_{G}(\lambda_{G}^{\ell})$ and $\mu_{B}^{\ell}=\mu_{B}(\lambda_{B}^{\ell})$, where 
\begin{equation}
\mu_G(\lambda):=1-\frac{\delta}{2\delta+\psi(\lambda)\alpha}\text{ and }\mu_B(\lambda):=\frac{\delta}{2\delta+\psi(\lambda)\alpha}. \label{eq:belief}
\end{equation}
For the corner case where $\lambda_G^\ell=\lambda_B^\ell=0$, 
we assume $\mu_{G}^{\ell}=\mu_{B}^{\ell}=1/2$, which are the limits of $\mu_{G}(\lambda)$ and $\mu_{B}(\lambda)$ as $\lambda\to 0$.

Since $\mu_{j}^{\ell}$ depends on $\lambda_j^\ell$, we can express buyers' expected search payoffs as
\begin{equation}
    \begin{split}
    u_G(\lambda):=&\phi(\lambda)(\mu_G(\lambda)u_H+(1-\mu_G(\lambda))u_L-p),\\
    u_B(\lambda):=&\phi(\lambda)(\mu_B(\lambda)u_H+(1-\mu_B(\lambda))u_L-p).
\end{split}\label{eq:payoff}
\end{equation}

The following lemma reports a crucial difference between $u_{G}$ and $u_{B}$. 

\begin{lem}\label{lem:symmetric_uj_behavior}
Suppose $(u_{H}+u_{L})/2>p$. Then, $u_{B}(\lambda)$ is always decreasing, while $u_{G}(\lambda)$ is monotone decreasing if and only if 
\begin{equation*}
k\leq\underline{k}:=\frac{1+\sqrt{1-\frac{u_{H}-u_{L}}{2(u_{H}-p)}}}{2}.	
\end{equation*}
If $k>\underline{k}$, then there exist $\overline{\lambda}_{G}>\underline{\lambda}_{G}>0$ such that $u_{G}(\lambda)$ is increasing if and only if $\lambda\in(\underline{\lambda}_{G},\overline{\lambda}_{G})$. 
\end{lem}
\begin{proof} 
The function $u_{B}(\lambda)$ is monotone decreasing because both $\phi(\lambda)$ and $\mu_{B}(\lambda)$ are decreasing. For $u_{G}(\lambda)$,
observe that $u_{G}^{\prime}(\lambda)$ has the same sign as 
\begin{eqnarray*}
h(\lambda):=-(1-k) 
   \left[\delta  \left(u_L-u_H\right)\left(2 \delta +\alpha  \lambda ^k\right)+(u_H-p)\left(2 \delta +\alpha  \lambda ^k\right)^2\right]+\alpha  \delta  k \lambda ^k
   \left(u_H-u_L\right).
\end{eqnarray*}
This is a quadratic equation of $\lambda^k$, whose maximal value is 
\begin{align*}
\frac{(u_{H}-u_{L}-4(1-k)(u_{H}-p))^{2}\delta^{2}}{4(1-k)(u_{H}-p)}-2(1-k)\delta^{2}(u_{H}+u_{L}-2p).
\end{align*}
The cutoff $\underline{k}\in(0,1)$ equates this maximal value to $0$.
It follows that if $k\leq\underline{k}$ then $h(\lambda)\leq 0$ for any $\lambda\in\mathbb{R}_+$, so $u_{G}(\lambda)$ is monotone decreasing. 

Suppose $k>\underline{k}$. Then, $h(\lambda)=0$ has two solutions, $\underline{\lambda}_{G}$ and $\overline{\lambda}_{G}(>\underline{\lambda}_{G})$. We conclude by showing $\underline{\lambda}_{G}>0$, which holds if and only if $h(0)<0$ and $h^{\prime}(0)>0$. Observe that $h(0)=-2(1-k)\delta^2(u_H+u_L-2p)<0$ and 
\begin{align*}
    h'(0)=\alpha\delta(u_{H}-u_{L}-4(1-k)(u_{H}-p))>0\iff k>\underline{k}^{\prime}:=1-\frac{u_{H}-u_{L}}{4(u_{H}-p)}.
\end{align*}
Since $\underline{k}^{\prime}\leq\underline{k}\Leftrightarrow (u_{H}+u_{L})/2\geq p$, this inequality holds whenever $k>\underline{k}$. \end{proof}

To understand this result, recall that $\lambda$ is the ratio of buyers to sellers in a submarket. Its increase has a direct negative \emph{congestion effect} on buyers, because their matching rate $\phi(\lambda)$ decreases in $\lambda$. 
In our model, however, it has an additional \emph{informational effect}: With relatively more buyers, sellers trade faster, updating their ratings more frequently. This implies a lower belief for rating $B$ and higher belief for rating $G$. Since buyers' expected payoffs are always increasing in $\mu_{j}^{\ell}$,  this informational effect works in the \emph{same} direction as the congestion effect in submarket $(B,\ell)$ and in the \emph{opposite} direction in submarket $(G,\ell)$. 

If $k\leq \underline{k}$,  the congestion effect always outweighs the informational effect, so $u_{G}$ is monotone decreasing. Intuitively, $k$ represents the output elasticity of buyers for match formation (see \cref{fn:Cobb_Douglas}). Therefore, for $k$ small, the total number of matches is not sensitive to the measure of buyers; but this means that the congestion effect is strong from an individual buyer's perspective, because a buyer's match rate is the total match rate divided by the measure of buyers. Consequently, $u_{G}(\lambda)$ is influenced  relatively less by the informational effect.  



However, if $k>\underline{k}$ and $\lambda\in(\underline{\lambda}_{G},\overline{\lambda}_{G})$, then $u_G(\lambda)$ is increasing in $\lambda$. In this case, the market exhibits a \emph{feedback loop} among G-rated sellers: (i) More frequent trading leads to 
higher perceived quality, while (ii) higher perceived quality attracts more buyers. The first is purely the informational effect. The second occurs only when the informational effect dominates the congestion effect. As will be clear shortly, the feedback loop is the key factor that leads to statistical discrimination.

\subsection{Trade vs. No Trade}\label{subsec:trade_cond}

For a preliminary analysis, we characterize the condition for trade to occur in a steady state. 

\begin{prop}\label{prop:trade_vs_no_trade}
    If $(u_{H}+u_{L})/2\leq p$ then there is no trade in steady state (i.e., $\forall j,\ell$, either $Q_j^\ell=0$ or $P_{Hj}^{\ell}+P_{Lj}^{\ell}=0$). Otherwise, there must be trade in all submarkets (i.e., $\forall j,\ell$, $\lambda_j^\ell>0$). 
\end{prop}
\begin{proof}
Suppose $(u_{H}+u_{L})/2\leq p$. Since $\mu_{B}^{\ell}\leq 1/2\leq \mu_{G}^{\ell}$ for all $(\lambda_{B}^{\ell},\lambda_{G}^{\ell})$, there are three cases to consider: (i) $\lambda_{B}^{\ell}>0$, (ii) $\lambda_{B}^{\ell}=0<\lambda_{G}^{\ell}$, and (iii) $\lambda_{G}^{\ell}=0$. In case (i), $\mu_{B}^{\ell}<1/2$ yielding $u_{B}(\lambda_{B}^{\ell})<0$, which contradicts $\lambda_{B}^{\ell}>0$. In case (ii), 
setting $\psi(\lambda_B^\ell)=0$ and $\psi(\lambda_G^\ell)>0$ in \eqref{eq:stationary} implies $P_{HG}^{\ell}+P_{LG}^{\ell}=0$, i.e., there are no $G$-rated sellers. Together with $\lambda_{B}^{\ell}=0$, this implies that there is no trade. In case (iii), clearly, there is no trade. 

Now suppose $(u_{H}+u_{L})/2>p$. If $\lambda_{j}^{\ell}=0$, then $u_{j}(0)=\phi(0)((u_{H}+u_{L})/2-p)$ is arbitrarily large, which is a contradiction. Therefore, $\lambda_{j}^{\ell}>0$ for all $j,\ell$.
\end{proof} 
Since the no-trade case is trivial, we maintain the following assumption hereafter: 
\begin{ass}\label{as:always_trade}
$(u_{H}+u_{L})/2>p$, and thus $\forall j,\ell,\ \lambda_{j}^{\ell}>0$ in equilibrium. 
\end{ass}

\subsection{Non-discriminatory equilibrium}\label{ss:symmetric_equil_characterize} 

We first characterize non-discriminatory equilibria. By definition, they do not depend on group identity $\ell$, so we focus on characterizing the total measures of consumers searching for rating $j=G,B$, denoted by $Q_{j}(:=Q_{j}^{1}+Q_{j}^{2})$, and the corresponding group-blind queue lengths, denoted by $\lambda_{j}$. 

The equilibrium tuple $(\lambda_{G},\lambda_{B})$ should satisfy two conditions.  First, market clearing requires: 
\begin{align}\label{eq:symmetric_market_clear}
    \begin{split}
    Q = \Theta(\lambda_{G},\lambda_{B}) :=&~ \lambda_{G}\sum_{i,\ell}P_{iG}^{\ell}+\lambda_{B}\sum_{i,\ell}P_{iB}^{\ell}\\
    =&~ \frac{\lambda_{G}\psi(\lambda_{B})(2\delta+\alpha\psi(\lambda_{G}))+\lambda_{B}\psi(\lambda_{G})(2\delta+\alpha\psi(\lambda_{B}))}{2(\delta\psi(\lambda_{G})+\delta\psi(\lambda_B)+\alpha\psi(\lambda_{G})\psi(\lambda_{B}))}.
    \end{split}
\end{align}
The function $\Theta$ specifies the total measure of buyers necessary to support queue lengths $\lambda_G$ and $\lambda_B$ in steady state (see \eqref{eq:stationary} and \cref{subsec:preliminaries}). Then, \eqref{eq:symmetric_market_clear} requires that ``demand'' of buyers to equal the corresponding ``supply''---the total measure $Q$ of buyers. 
Let $\lambda_{B}^{MC}(\lambda_{G})$ be the implicit function defined by \eqref{eq:symmetric_market_clear}, where superscript $MC$ stands for ``market clearing.'' As depicted in \cref{fig:symmetric_equil}, $\lambda_{B}^{MC}(\lambda_{G})$ is monotone decreasing,  $\lim\limits_{\lambda_{G}\to 0}\lambda_{B}^{MC}(\lambda_{G})=\infty$, and $\lim\limits_{\lambda_{G}\to \infty}\lambda_{B}^{MC}(\lambda_{G})=0$.\footnote{The first property holds because $\Theta(\lambda_G,\lambda_B)$ strictly increases in both arguments. The last two are because, under \cref{as:always_trade}, the weighted average of $\lambda_{G}$ and $\lambda_{B}$ should be $Q$.} Intuitively, if more buyers attend the $G$-rated submarket, then for the whole market to clear, fewer buyers must search in the $B$-rated submarket.

 Second,  buyers must be indifferent between the two submarkets: 
    \begin{equation}\label{eq:symmetric_equal_profit}
    u_{G}(\lambda_{G})=u_{B}(\lambda_{B}). 
    \end{equation}
If $(\lambda_{G},\lambda_{B})$ satisfies \eqref{eq:symmetric_equal_profit}, then $\lambda_{G}\geq \lambda_{B}$, which follows from $\mu_{G}\geq 1/2\geq \mu_{B}$ and so $\phi(\lambda_{G})\leq \phi(\lambda_{B})$.
Similarly to $\lambda_{B}^{MC}(\lambda_{G})$, let $\lambda_{B}^{BI}(\lambda_{G})$ be the implicit function defined by \eqref{eq:symmetric_equal_profit}, where superscript $BI$ stands for ``buyer indifference.''  It is straightforward to show that, as shown in \cref{fig:symmetric_equil}, $\lim_{\lambda_{G}\to 0}\lambda_{B}^{BI}(\lambda_{G})=0$, $\lim_{\lambda_{G}\to\infty}\lambda_{B}^{BI}(\lambda_{G})>0$, and $\lambda_{B}^{BI}(\lambda_{G})$ is continuous. Recall from \Cref{lem:symmetric_uj_behavior} that $u_B(\cdot)$ is monotone decreasing, but $u_G(\cdot)$ may or may not be, depending on the relative strength of the congestion and informational effects. When $u_G(\cdot)$ is monotone decreasing, $\lambda_{B}^{BI}(\cdot)$ is monotone increasing, but when $u_G(\cdot)$ is non-monotone, so is $\lambda_{B}^{BI}(\cdot)$. 

\begin{figure}[htbp]
\centering{}\beginpgfgraphicnamed{fig:symmetric_equil} \begin{tikzpicture}[scale=1]

\draw[line width=0.5pt] (0, 5) -- (0,0) -- (6,0);

\fill (0,5) node[left] {\footnotesize{$\lambda_{B}$}};
\fill (6,0) node[below] {\footnotesize{$\lambda_{G}$}}; 
\fill (0,0) node[below] {\footnotesize{$0$}};


\draw[blue,line width=0.5pt] plot[smooth] (0,0)--(0.0032,0.0395)--(0.0988,0.9791)--(0.1945,1.6351)--(0.2901,2.1292)--(0.32,2.2596)--(0.4815,2.8255)--(0.5771,3.0781)--(0.6368,3.2127)--(0.8641,3.6062)--(0.9537,3.7220)--(1.1511,3.9226)--(1.2705,4.0153)--(1.3424,4.0630)--(1.5873,4.1886)--(1.9042,4.2905)--(2.2210,4.3478)--(2.5378,4.3762)--(2.8547,4.3858)--(3.1715,4.3830)--(3.4883,4.3720)--(3.8052,4.3558)--(4.1220,4.3360)--(4.4388,4.3145)--(4.7557,4.2917)--(5.0725,4.2689)--(5.3894,4.2461)--(5.7062,4.2236)--(6.0230,4.2018);

\fill (5,4.27) node[above] {\footnotesize{$\lambda_{B}^{BI}(\lambda_{G})$}}; 

\draw[red,dashed,line width=0.5pt] plot[smooth] (0.2516,5.4006)--(0.3013,4.2226)--(0.3354,3.7490)--(0.4007,3.1781)--(0.5001,2.6951)--(0.5995,2.4113)--(0.6492,2.3078)--(0.6676,2.2739)--(0.9999,1.8868)--(1.3321,1.6864)--(1.6644,1.5550)--(1.9966,1.4581)--(2.3289,1.3814)--(2.6611,1.3179)--(2.9934,1.2637)--(3.3256,1.2163)--(3.6578,1.1743)--(3.9901,1.1364)--(4.3223,1.1020)--(4.6546,1.0705)--(4.9868,1.0414)--(5.3191,1.0143)--(5.6513,0.9890)--(5.9835,0.9654);

\fill (5,1.05) node[above] {\footnotesize{$\lambda_{B}^{MC}(\lambda_{G})$}}; 

\fill (canvas cs:x=0.4504cm,y=2.7320cm) circle (2pt);

\draw[dotted,line width=0.5pt] (0.4504,0)--(0.4504,2.732)--(0,2.732);

\begin{scope}[xshift=7.5cm]
\draw[line width=0.5pt] (0, 5) -- (0,0) -- (6,0);

\fill (0,5) node[left] {\footnotesize{$\lambda_{B}$}};
\fill (6,0) node[below] {\footnotesize{$\lambda_{G}$}};
\fill (0,0) node[below] {\footnotesize{$0$}};


\draw[blue,line width=0.5pt] plot[smooth] (0,0)--(0.0008,0.0590)--(0.0257,1.4311)--(0.0506,2.2115)--(0.0755,2.7029)--(0.1004,3.0236)--(0.1253,3.2344)--(0.1502,3.3704)--(0.1752,3.4537)--(0.2001,3.4992)--(0.2250,3.5169)--(0.5333,2.9937)--(0.8417,2.3974)--(1.1500,1.9785)--(1.4583,1.6899)--(1.7667,1.4861)--(2.0750,1.3388)--(2.3833,1.2279)--(2.6917,1.1445)--(3,1.0788)--(3.3333,1.0240)--(3.6667,0.9797)--(4,0.9456)--(4.3333,0.9187)--(4.6667,0.8979)--(5,0.8800)--(5.3333,0.8661)--(5.6667,0.8545)--(6,0.8448);

\fill (1,2.5) node[right] {\footnotesize{$\lambda_{B}^{BI}(\lambda_{G})$}};

\fill (0.1,4.5) node[right] {\footnotesize{$\lambda_{B}^{MC}(\lambda_{G})$}};

\draw[red,dashed,line width=0.5pt] plot[smooth] (0.0943,4.8556)--(0.1153,4.2181)--(0.1362,3.8311)--(0.1571,3.5671)--(0.178,3.3727)--(0.1989,3.2218)--(0.2199,3.0999)--(0.2408,2.9985)--(0.2617,2.9121)--(0.2826,2.8371)--(0.3036,2.7709)--(0.3245,2.7118)--(0.3454,2.6584)--(0.3513,2.6442)--(0.6651,2.1716)--(0.9789,1.9121)--(1.2928,1.7278)--(1.6066,1.5839)--(1.9204,1.4660)--(2.2342,1.3665)--(2.5480,1.2808)--(2.8618,1.2059)--(3.1757,1.1398)--(3.4895,1.0808)--(3.8033,1.0277)--(4.1171,0.9797)--(4.4309,0.9361)--(4.7447,0.8962)--(5.0586,0.8595)--(5.3724,0.8258)--(5.6862,0.7946)--(6,0.7656);

\fill (canvas cs:x=0.17cm,y=3.4398cm) circle (2pt);
\fill (canvas cs:x=1.7667cm,y=1.4862cm) circle (2pt);
\fill (canvas cs:x=4.6667cm,y=0.8976cm) circle (2pt);

\draw[dotted,line width=0.5pt] (0.17,0)--(0.17,3.4398);
\draw[dotted,line width=0.5pt] (1.7667,0)--(1.7667,1.4862);
\draw[dotted,line width=0.5pt] (4.6667,0)--(4.6667,0.8976);     
\end{scope}

\end{tikzpicture} \endpgfgraphicnamed\caption{\label{fig:symmetric_equil} The blue solid curves depict the buyer-indifference condition \eqref{eq:symmetric_equal_profit}, while the red dashed curves depict the market clearing condition \eqref{eq:symmetric_market_clear}. The common parameters used for this figure are $\delta=1$, $\alpha=0.1$, $u_{H}=2$, and $u_{L}=p=1$ 
}
\end{figure}

Given the two functions defined above, equilibrium characterization reduces to finding $\lambda_{G}$ such that $\lambda_{B}^{MC}(\lambda_{G})=\lambda_{B}^{BI}(\lambda_{G})$. Equilibrium existence follows immediately since both functions are continuous, and $\lambda_{B}^{MC}(\lambda_{G})>\lambda_{B}^{BI}(\lambda_{G})$, for $\lambda_{G}$ close to $0$, while $\lambda_{B}^{MC}(\lambda_{G})<\lambda_{B}^{BI}(\lambda_{G})$, for $\lambda_{G}$ sufficiently large.  

In general, these two curves, $\lambda_{B}^{MC}$ and $\lambda_{B}^{BI}$, may have multiple crossings (see  the right panel of \cref{fig:symmetric_equil}), in which case multiple equilibria exist.\footnote{Multiple equilibria can be ranked in terms of buyer surplus as equilibrium buyer payoff is given by $u_B(\lambda_B)$. Hence, larger equilibrium $\lambda_G$ corresponds to smaller $\lambda_B=\lambda_B^{MC}(\lambda_G)$ and higher buyer surplus.
}
This multiplicity may arise because, while $\lambda_{B}^{MC}(\cdot)$ is always decreasing, $\lambda_{B}^{BI}(\cdot)$ may not be monotone increasing, which ``mirrors'' the non-monotonicity of $u_{G}(\cdot)$ reported in \cref{lem:symmetric_uj_behavior}. 

The above argument suggests that if $u_{G}(\cdot)$---or equivalently, $\lambda_{B}^{BI}(\cdot)$---is monotone, then there exists a unique non-discriminatory equilibrium. In fact, this uniqueness can be strengthened in the following fashion. 

\begin{prop}\label{prop:monotone_unique}
    If $k\leq\underline{k}$, then there exists a unique equilibrium, which is non-discriminatory.
\end{prop}
\begin{proof}
    It suffices to rule out discriminatory equilibria. When $(u_{H}+u_{L})/2>p$, buyers must be indifferent over all submarkets:
\begin{equation*}
u_{G}(\lambda_{G}^{1})=u_{B}(\lambda_{B}^{1})=u_{B}(\lambda_{B}^{2})	=u_{G}(\lambda_{G}^{2}).
\end{equation*}
Since both $u_{G}$ and $u_{B}$ are monotone, it must be that $\lambda_{G}^{1}=\lambda_{G}^{2}$ and $\lambda_{B}^{1}=\lambda_{B}^{2}$. 
\end{proof}

\subsection{Discriminatory Equilibrium}\label{s:asymmetric_equil} 

We now investigate discriminatory equilibria in which buyers condition their search on sellers' {\it group identities} as well as their ratings. We derive a condition under which such equilibria exist and then compare them to non-discriminatory equilibria studied in \cref{ss:symmetric_equil_characterize}. 

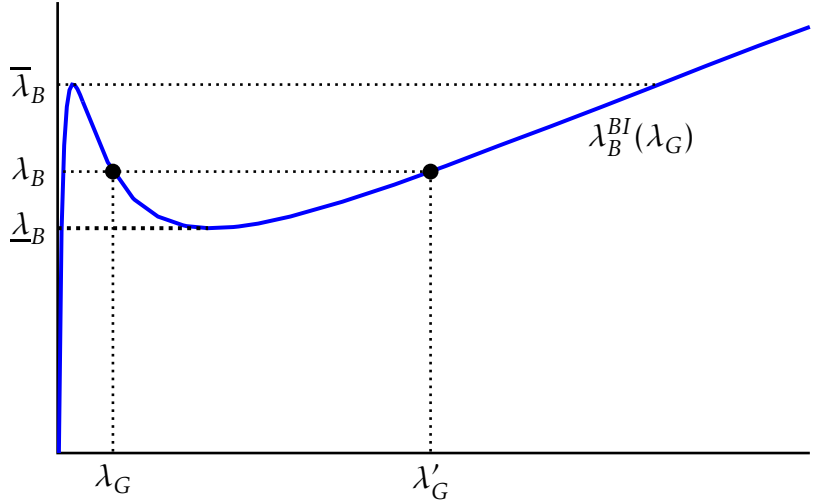
\begin{figure}[htbp]
\centering{}\beginpgfgraphicnamed{fig:asymmetric_equil} \begin{tikzpicture}[scale=0.9]

\draw[line width=0.5pt] (0, 6) -- (0,0) -- (10,0);



\draw[blue,line width=0.5pt] plot[smooth] (0.0167,0.0010)--(0.0519,2.9444)--(0.0870,4.0944)--(0.1222,4.6074)--(0.1574,4.8292)--(0.1926,4.9024)--(0.2278,4.8948)--(0.2630,4.8429)--(0.2981,4.7667)--(0.3333,4.6776);

\draw[blue,line width=0.5pt] plot[smooth] (0.3333,4.6776)--(0.6667,3.8676)--(1,3.3930)--(1.0238,3.3697)--(1.3333,3.1464)--(1.6667,3.0300)--(1.7143,3.0189)--(2,2.9916)--(2.3333,3.0066)--(2.4048,3.0141)--(2.6667,3.0528)--(3.0952,3.1470)--(3.7857,3.3438)--(4.4762,3.5709)--(5.1667,3.8214)--(5.8571,4.0858)--(6.5476,4.3436)--(7.2381,4.6080)--(7.9286,4.8773)--(8.6190,5.1501)--(9.3095,5.4154)--(10,5.6700);

\draw[dotted,line width=0.5pt] (0,4.9024)--(8,4.9024);
\fill (0,4.9024) node[left] {\footnotesize{$\overline{\lambda}_{B}$}};

\draw[dotted,line width=0.5pt] (0,2.9916)--(2,2.9916);
\fill (0,2.9916) node[left] {\footnotesize{$\underline{\lambda}_{B}$}};

\fill (canvas cs:x=0.7351cm,y=3.7447cm) circle (2pt);

\fill (canvas cs:x=4.9580cm,y=3.7447cm) circle (2pt);
    
\draw[dotted,line width=0.5pt] (0.7351,3.7447)--(0.7351,0);

\fill (0.7351,0) node[below] {\footnotesize{$\lambda_{G}^1$}};

\draw[dotted,line width=0.5pt] (4.9580,3.7447)--(4.9580,0);

\fill (4.9580,0) node[below] {\footnotesize{$\lambda_{G}^2$}};

\draw[dotted,line width=0.5pt] (4.9580,3.7447)--(0,3.7447);
\fill (0,3.7447) node[left] {\footnotesize{$\lambda_{B}$}}; 

\fill (7.75,4.6080) node[below] {\footnotesize{$\lambda_{B}^{BI}(\lambda_{G})$}}; 
    
\end{tikzpicture} \endpgfgraphicnamed\caption{\label{fig:asymmetric_equil} The blue solid curve represents the buyer-indifference condition \eqref{eq:symmetric_equal_profit}. The parameters used for this figure are $\delta=0.2$, $\alpha=0.5$, $u_{H}=3$, $u_{L}=1$, $w=1.5$
.}
\end{figure}

We first explain how the non-monotonicity of $u_{G}(\lambda)$ leads to the existence of discriminatory equilibria. Suppose $u_{G}(\lambda)$ is non-monotone, and fix $\lambda_{B}$ such that $u_{B}(\lambda_{B})=u_{G}(\lambda_{G}^{1})=u_{G}(\lambda_{G}^{2})$ for two distinct values: 
in \cref{fig:asymmetric_equil}, this means fixing any  $\lambda_{B}\in[\underline{\lambda}_{B},\overline{\lambda}_{B}]$. Recall from \eqref{eq:symmetric_market_clear} that $\Theta(\lambda_{G}^{\ell},\lambda_{B})$ is the measure of buyers that supports $(\lambda_{G}^{\ell},\lambda_{B})$ as a non-discriminatory equilibrium. Now, let $Q=(\Theta(\lambda_G^1,\lambda_B)+\Theta(\lambda_G^2,\lambda_B))/2$.

Consider a steady state with queue lengths $\lambda_G^1,\lambda_G^2$ and $\lambda_{B}^{1}=\lambda_{B}^{2}=\lambda_{B}$. By construction, buyers are indifferent over all submarkets: 
\begin{equation*}
u_{B}(\lambda_{B}^{1})=u_{B}(\lambda_{B}^{2})=u_{G}(\lambda_{G}^{1})=u_{G}(\lambda_{G}^{2}).
\end{equation*}
In addition, $Q$ is chosen to satisfy market clearing. Therefore, this (discriminatory) strategy profile is a steady-state equilibrium. 


 The mechanism behind our discriminatory equilibria is the novel feedback loop between ratings and trade. More buyers search for G1 sellers than G2 sellers because, despite the same G-rating,  buyers perceive the former to be more favorable than the latter  (i.e., $\mu_{G}^{1}>\mu_{G}^{2}$).  This
 causes the former sellers to be
   hired and reviewed more frequently than the latter sellers, making the G1 more accurate, and thus more favorable, than the G2 rating. In turn, this justifies buyers' discriminatory search behavior (i.e., $\lambda_{G}^{1}>\lambda_{G}^{2}$), completing the loop. 

The same feedback loop does not exist for $B$-rated sellers. For them, the informational effect and the congestion effect reinforce each other, in that more congestion not only decreases buyers' matching rate but also lowers the average quality of sellers; 
that is, if $\lambda_{B}^{1}<\lambda_{B}^{2}$ then $\phi(\lambda_{B}^{1})>\phi(\lambda_{B}^{2})$ and $\mu_{B}(\lambda_{B}^{1})<\mu_{B}(\lambda_{B}^{2})$, both of which contribute to $u_{B}(\lambda_{B}^{1})>u_{B}(\lambda_{B}^{2})$. In other words, increased attention toward one group self-corrects.  Hence, $\lambda_B^1=\lambda_B^2$ necessarily holds in equilibrium to ensure buyers' indifference between $(B,1)$ and $(B,2)$ submarkets. 

The above difference between $G$ and $B$ ratings suggests that our ratings-guided market features ``glass ceiling''---discrimination concentrated at the top end of the market. Statistical discrimination, if it occurs, hurts only good-rated sellers from the disadvantageous group but not bad-rated sellers. 

Clearly, the feedback loop occurs only when $u_{G}(\lambda)$ is non-monotone. The following result shows that whenever $u_{G}(\lambda)$ is non-monotone, there is a range of $Q$'s that produce discriminatory equilibria. 

\begin{thm}\label{thm:discriminatory_existence} 
If $k>\underline{k}$, then there exist $0<\underline{Q}<\overline{Q}$ such that a discriminatory equilibrium exists if $Q\in(\underline{Q},\overline{Q})$ and only if $Q\in[\underline{Q},\overline{Q}]$.
\end{thm}

\begin{proof}
If $k\in(\underline{k},1)$ then, as shown in \cref{lem:symmetric_uj_behavior}, there exist $\overline{\lambda}_{G}>\underline{\lambda}_{G}>0$ such that $u_{G}(\lambda)$ is increasing if and only if $\lambda\in(\underline{\lambda}_{G},\overline{\lambda}_{G})$. Let $\overline{\lambda}_{B}$ and $\underline{\lambda}_{B}$ be the values such that $\overline{\lambda}_{B}=\lambda_{B}^{BI}(\underline{\lambda}_{G})$  (i.e., $u_{B}(\overline{\lambda}_{B})=u_{G}(\underline{\lambda}_{G})$) and $\underline{\lambda}_{B}=\lambda_{B}^{BI}(\overline{\lambda}_{G})$ (i.e., $u_{B}(\underline{\lambda}_{B})=u_{G}(\overline{\lambda}_{G})$), respectively (see \cref{fig:asymmetric_equil}). Then, for each $\lambda_{B}\in[\underline{\lambda}_{B},\overline{\lambda}_{B}]$, the inverse of $\lambda_B^{BI}$ is a correspondence, denoted by $\lambda_B^{BI}{}^{-1}(\lambda)=\{h_1(\lambda),h_2(\lambda),h_3(\lambda)\}$, where $h_{1}(\lambda)\leq h_{2}(\lambda)\leq h_{3}(\lambda)$, with at least one inequality holding strictly if $\lambda=\underline{\lambda}_{B}$ or $\lambda=\overline{\lambda}_{B}$ and both inequalities holding strictly otherwise. By construction, all $h_m$'s are continuous over $(\underline{\lambda}_{B},\overline{\lambda}_{B})$. In addition, $h_{1}$ and $h_{3}$ are strictly increasing, while $h_{2}$ is strictly decreasing.

As explained above, every discriminatory equilibrium can be spanned by two distinct points, $\lambda_{G}^{1}$ and $\lambda_{G}^{2}$, that correspond to the same value of $\lambda_{B}\in[\underline{\lambda}_{B},\overline{\lambda}_{B}]$. Combining this with the fact that for each $\lambda_{B}\in(\underline{\lambda}_{B},\overline{\lambda}_{B})$, there are three such values, $h_{1}(\lambda_{B})$, $h_{2}(\lambda_{B})$, and $h_{3}(\lambda_{B})$, it follows that 
the set of all $Q$'s that are compatible with discriminatory equilibria consists of the following three subsets:\footnote{We exclude $\overline{\lambda}_B$ and $\underline{\lambda}_B$ when defining $I_1$ and $I_2$, respectively, because $h_1(\overline{\lambda}_B)=h_2(\overline{\lambda}_B)$ and $h_2(\underline{\lambda}_B)=h_3(\underline{\lambda}_B)$.}
\begin{align*}
    I_1=&\left\{\frac{\Theta(h_1(\lambda_{B}),\lambda_{B})+\Theta(h_2(\lambda_{B}),\lambda_{B})}{2}\Big|\lambda_{B}\in[\underline{\lambda}_{B},\overline{\lambda}_{B})\right\};\\
    I_2=&\left\{\frac{\Theta(h_2(\lambda_{B}),\lambda_{B})+\Theta(h_3(\lambda_{B}),\lambda_{B})}{2}\Big|\lambda_{B}\in(\underline{\lambda}_{B},\overline{\lambda}_{B}]\right\};\\
    I_3=&\left\{\frac{\Theta(h_1(\lambda_{B}),\lambda_{B})+\Theta(h_3(\lambda_{B}),\lambda_{B})}{2}\Big|\lambda_{B}\in[\underline{\lambda}_{B},\overline{\lambda}_{B}]\right\}.
\end{align*}
Since $\Theta$ and $h_m$ are continuous, each $I_m$ is an interval. Furthermore, the left (right) boundary of $I_3$ is contained in $I_1$ ($I_2$) since $h_1(\overline{\lambda}_B)=h_2(\overline{\lambda}_B)$ and $h_2(\underline{\lambda}_B)=h_3(\underline{\lambda}_B)$. Therefore, $I=\cup I_m$ is an interval. 
\end{proof}

To see why a discriminatory equilibrium does not exist when $Q$ is sufficiently small or sufficiently large, notice that
the informational effect is bounded as the belief $\mu_G$ is restricted to $[0.5,1]$, so the second term for $u_{G}(\lambda_{G})$ in \eqref{eq:payoff} varies only between $(u_{H}+u_{L})/2-p$ and $u_{H}-p$. Meanwhile, the congestion effect is unbounded as the matching rate $\phi(\lambda_{B})$ ranges from 0 to $\infty$. Consequently, when the market is extremely sparse or crowded, the congestion effect dominates the informational effect, preventing the feedback loop from occurring. 

\subsection{Stability of Ratings-Based Discrimination}

The analysis so far has demonstrated that non-discriminatory and discriminatory equilibria can coexist. The possibility of discriminatory equilibria is intriguing, but they would not be worrisome if they were not robust phenomena. We now examine whether (and when) discriminatory equilibria are stable.


For tractability, hereafter, we restrict attention to the parameter space that satisfies the following assumption.\footnote{\cref{as:unique_symmetric} permits the existence of discriminatory equilibria, e.g.,  when $k$ is above but close to $\underline{k}$. }
\begin{ass}\label{as:unique_symmetric}
For all $Q\in\mathbb{R}_{+}$, there exists a unique non-discriminatory equilibrium. 
\end{ass}



To provide a relevant stability concept, let $U(Q)$ denote buyers' expected payoff in the unique non-discriminatory equilibrium when the total measure of buyers is $Q$, holding all other parameters fixed. In other words, if $(\lambda_{B},\lambda_{G})$ is a non-discriminatory equilibrium with measure $Q$ of buyers, then $U(Q)=u_{B}(\lambda_{B})=u_{G}(\lambda_{G})$. \cref{as:unique_symmetric} ensures that $U(Q)$ is well-defined. 

A steady-state equilibrium can be summarized by a pair $(Q^{1},Q^{2})$ such that $(Q^{1}+Q^{2})/2=Q$ (market clearing) and $U(Q^{1})=U(Q^{2})$ (buyer indifference over all submarkets). An equilibrium is non-discriminatory if $Q^{1}=Q^{2}$ and  discriminatory if $Q^{1}\neq Q^{2}$. Based on this observation, we adopt the following tractable notion of stability. 

\begin{defi}\label{def:stable}
An equilibrium with $(Q^{1},Q^{2})$ is stable if $U^{\prime}(Q^{1})+U^{\prime}(Q^{2})\leq 0$ and unstable otherwise. 
	\label{defi:stable}
\end{defi}

\cref{defi:stable} captures the idea that a stable outcome should restore itself after a small perturbation. Specifically, suppose a small measure of buyers shift their searches from group $1$ to group $2$ sellers. It changes the expected payoff of buyers (still) searching for group $1$ sellers by $-U^{\prime}(Q^{1})$ and that of those targeting group $2$ sellers by $U^{\prime}(Q^{2})$. If $-U^{\prime}(Q^{1})\geq U^{\prime}(Q^{2})$ then those buyers who left group $1$ would move back, restoring the original equilibrium. Otherwise, even more buyers would switch to group 2, drifting the economy further away from the original equilibrium.

The following result provides stability conditions for both non-discriminatory and discriminatory equilibria.

\begin{thm}\label{thm:stable}${}$
    \begin{itemize}
        \item When $k\le \underline{k}$, the unique (non-discriminatory) equilibrium is stable.
        \item When $k>\underline{k}$, let $\underline{Q}$ and $\overline{Q}$ be as defined in \cref{thm:discriminatory_existence}. Then,
        \begin{itemize}
            \item there exists an interval $(\underline{Q}',\overline{Q}')\subset [\underline{Q},\overline{Q}]$ such that the non-discriminatory equilibrium is stable if and only if $Q\not\in(\underline{Q}',\overline{Q}')$; 
            \item there exists a stable  discriminatory equilibrium whenever $Q\in (\underline{Q},\overline{Q})$.
        \end{itemize}
    \end{itemize}
\end{thm}
\begin{proof}
In a non-discriminatory equilibrium, $Q^{1}=Q^{2}=Q$. Therefore, it is stable if and only if $U^{\prime}(Q)\leq 0$, which is equivalent to the equilibrium value of $\lambda_{B}$---the intersection of $\lambda_{B}^{MC}$ and $\lambda_{B}^{BI}$---increasing in $Q$. Since $\lambda_{B}^{MC}$ uniformly increases in $Q$, the required property necessarily holds when $\lambda_{B}^{BI}$ is increasing at the equilibrium value of $\lambda_{G}$. 

If $k\leq\underline{k}$ then $\lambda_{B}^{BI}$ is monotone increasing, so the (non-discriminatory) equilibrium is always stable. If $k>\underline{k}$ then $\lambda_{B}^{BI}$ is strictly decreasing if and only if 
 \begin{align*}
     Q\in(\underline{Q}',\overline{Q}'):=\left\{\Theta(h_2(\lambda_B),\lambda_B)\Big|\lambda_B\in (\underline{\lambda}_B,\overline{\lambda}_B) \right\}.
 \end{align*}
Since $h_1\le h_2\le h_3$, 
$\frac{\Theta(h_1(\lambda_B),\lambda_B)+\Theta(h_2(\lambda_B),\lambda_B)}{2}\le \Theta(h_2(\lambda_B),\lambda_B)\le\frac{\Theta(h_3(\lambda_B),\lambda_B)+\Theta(h_2(\lambda_B),\lambda_B)}{2}$; hence, $(\underline{Q}',\overline{Q}')\subset [\underline{Q},\overline{Q}]$. 

Suppose $k>\underline{k}$ and $Q\in (\underline{Q},\overline{Q})$, so that a discriminatory equilibrium exists. 
Consider the function $g:[0,Q)\to\mathcal{R}$ such that 
\begin{equation*}
g(x)=U\left(Q+x\right)-U\left(Q-x\right). 	
\end{equation*}
If $x$ is close to $Q$ then $g(x)<0$, because $U(2Q)<\infty$ while $\lim_{q\to 0}U(q)=\infty$. Meanwhile, the existence of discriminatory equilibrium implies that there exists $x^{\ast}\in(0,Q)$ such that $g(x^{\ast})=0$ (i.e., $x^{\ast}=|Q^{1}-Q^{2}|$). Combining these with continuity of $g$, it follows that there exists $x^{\ast\ast}\in[x^{\ast},Q)$ such that 
\begin{equation*}
g(x^{\ast\ast})=0\text{ and }g^{\prime}(x^{\ast\ast})\leq 0. 
\end{equation*}
The first condition implies that $(Q+x^{\ast\ast},Q-x^{\ast\ast})$ is a discriminatory equilibrium, while the second condition implies that the equilibrium is stable. 
\end{proof}

\cref{thm:stable} shows that the non-discriminatory equilibrium is stable whenever a discriminatory equilibrium does not exist (either $k\leq\underline{k}$ or $k>\underline{k}$ and $Q\notin[\underline{Q},\overline{Q})$). However, if a discriminatory equilibrium exists ($k>\underline{k}$ and $Q\in(\underline{Q},\overline{Q})$),  the non-discriminatory equilibrium may not be stable (when $Q\in(\underline{Q}^{\prime},\overline{Q}^{\prime})$), while there necessarily exists a stable discriminatory equilibrium (which may or may not be the original discriminatory equilibrium). This latter result suggests that when they coexist, discriminatory equilibria are more stable than non-discriminatory equilibria---a noteworthy result that is novel in the literature.\footnote{No such distinction exists between non-discriminatory and discriminatory equilibria in existing theories of discrimination (e.g., \cite{coate1993antidiscrimination}). Contemporary work by \cite{gu2020searchmodelstatisticaldiscrimination} report similar results in a random search model with occupational choice. 
}   In this sense, discrimination is a real concern in the ratings-guided markets. What underlies these results is that the congestion effect is a stabilizing force (reducing buyers' incentives to leave their submarkets), while the informational effect is a potential source of instability.

\section{Discussions}

\subsection{Rating Quality}\label{ss:rate_quality_dis_equil} 

With ratings becoming increasingly prevalent, the associated technology has improved, allowing users access to more accurate information about their counterparties.  Will such a technological advance bring more fairness by weakening the role of prejudice in decision-making, or can it actually worsen discrimination? To address this question, we measure the quality of rating by $\beta:=\alpha/\delta$; this is an appropriate measure because $\alpha$ quantifies how fast ratings get corrected, while $\delta$ measures how fast ratings become obsolete. The following result shows that the effect of increasing $\beta$ is ambiguous in general.\footnote{\citet{echenique2022rationally} report a similar non-monotonicity result in terms of attention costs. Their underlying mechanism for discrimination is based on \citet{coate1993antidiscrimination} and distinct from ours.} 


\begin{prop}\label{prop:alpha:discrminatory}
If $k>\underline{k}$, 
then there exist $0<\underline{\beta}<\overline{\beta}$ such that a discriminatory equilibrium exists if $\beta\in(\underline{\beta},\overline{\beta})$ and only if $\beta\in[\underline{\beta},\overline{\beta}]$. 
\end{prop}
\begin{proof}  
We prove that if there is an equilibrium (whether discriminatory or not) with $Q$ and $\beta$, then effectively the same equilibrium exists with $Q\beta^{1/k}$ and $1$ as well, and vice versa. 
Let $\lambda_{j}^{\prime}:=\lambda_{j} \beta^{1/k}$. Then, the two equilibrium equations, \eqref{eq:symmetric_market_clear} and \eqref{eq:symmetric_equal_profit}, can be written as follows: 
\begin{equation*}
Q\beta^{1/k}=\frac{\lambda_{G}^{\prime}\psi(\lambda_B^{\prime})(2+\psi(\lambda_G^{\prime}))+\lambda_{B}^{\prime}\psi(\lambda_G^{\prime})(2+\psi(\lambda_B^{\prime}))}{4(\psi(\lambda_G^{\prime})+\psi(\lambda_B^{\prime})+\psi(\lambda_G^{\prime})\psi(\lambda_B^{\prime}))},
\end{equation*}
and 
\begin{equation*}
\phi(\lambda_{G}^{\prime})\left[\frac{1+\psi(\lambda_G^{\prime})}{2+\psi(\lambda_G^{\prime})}(u_{H}-u_{L})+u_{L}-p\right]=\phi(\lambda_{B}^{\prime})\left[\frac{1}{2+\psi(\lambda_B^{\prime})}(u_{H}-u_{L})+u_{L}-p)\right]. 
\end{equation*}
This implies that $(\lambda_{G}^{\prime},\lambda_{B}^{\prime})$ yields an equilibrium that is equivalent to the case where the measure of buyers is $Q\beta^{1/k}$ and $\alpha=\delta$. 
Together with \cref{thm:discriminatory_existence}, this implies that for a fixed value of $Q$, a discriminatory equilibrium exists if $\beta\in(\underline{\beta},\overline{\beta})$ and only if $\beta\in[\underline{\beta},\overline{\beta}]$, where $Q\underline{\beta}^{1/k}=\underline{Q}$ and $Q\overline{\beta}^{1/k}=\overline{Q}$. 
\end{proof}


For intuition, note that in our model, information is generated in two steps: (i) match formation and (ii) rating update. In particular, increasing sellers' matching rates (with more buyers) in step (i) has the same effect on the distribution of sellers' ratings as raising the quality of rating in step (ii). \cref{prop:alpha:discrminatory}---namely, that discrimination arises when the quality of rating belongs to an intermediate range---then follows from \cref{thm:discriminatory_existence}. This equivalence implies that an analogy of \cref{thm:stable} also holds for rating quality, as stated in the following result. 


\begin{prop}\label{prop:alpha:stable}
If $k>\underline{k}$ then there exists an interval $(\underline{\beta}',\overline{\beta}')\subset [\underline{\beta},\overline{\beta}]$ such that the non-discriminatory equilibrium is stable if and only if $\beta\not\in (\underline{\beta}',\overline{\beta}')$. 
\end{prop}

\cref{prop:alpha:discrminatory,prop:alpha:stable} together have the following implications: Raising $\beta$ could produce stable discriminatory equilibria (if above $\underline{\beta}$) and dis-stabilize non-discriminatory equilibria (if above $\underline{\beta}^{\prime}$). Further raising $\beta$, however, could stabilize non-discriminatory equilibria (if above $\overline{\beta}^{\prime}$) and eliminate discrimination in the market (if above $\overline{\beta}$).


\subsection{Asymmetric Group Sizes}\label{ssec:asymmetry}

Next, we extend the model to introduce asymmetric group sizes.  We assume that group $1$ is of size $\frac{1+\eta}{2}$ and group $2$ is of size $\frac{1-\eta}{2}$, where $\eta\in(0,1)$. 
We refer to group 1 as the ``majority'' and group 2 as the ``minority.'' We say that a discriminatory equilibrium is against group $\ell$, if $\lambda_G^{\ell}\le \lambda_G^{-\ell}$. In words, the discriminated group attracts relatively fewer buyers to $G$-rated sellers, and the market belief of their $G$-rated sellers is lower. 

\begin{thm}
	If $k\in(\underline{k},1)$ then for each $\eta\in(0,1)$, there exist $\underline{Q}_1<\underline{Q}_2<\overline{Q}_1<\overline{Q}_2$ s.t. a discriminatory equilibrium against type $\ell$ exists if $Q\in(\underline{Q}_{\ell},\overline{Q}_{\ell})$ and only if $Q\in[\underline{Q}_{\ell},\overline{Q}_{\ell}]$. Moreover, $\underline{Q}_1$ and $\overline{Q}_1$ are decreasing in $\eta$, while $\underline{Q}_2$ and $\overline{Q}_2$ are increasing in $\eta$.
	\label{thm:asymmetric:group}
\end{thm}
\begin{proof}
    Extending the arguments used in \cref{s:asymmetric_equil}, the following market clearing condition should hold: 
	\begin{align}
		Q=\frac{1+\eta}{2}\Theta(h_m(\lambda_B),\lambda_B)+\frac{1-\eta}{2}\Theta(h_{m'}(\lambda_B),\lambda_B).\label{eq:asymmetric}
	\end{align}
	If discrimination against the majority group 1 takes place, then $m<m'$. In addition, by the particular structure depicted in \cref{fig:asymmetric_equil}, $\underline{Q}_1$ is achieved by minimizing \eqref{eq:asymmetric} for $\lambda_B\in[\underline{\lambda}_B,\overline{\lambda}_B]$ with $m=1,m'=2$. $\overline{Q}_1$ is achieved by maximizing \eqref{eq:asymmetric} with $m=2,m'=3$.
 
    Similarly, if discrimination against the majority group 2 takes place, then $m>m'$.
    $\underline{Q}_2$ is achieved by minimizing \eqref{eq:asymmetric} with $m=2,m'=1$. $\overline{Q}_2$ is achieved by maximizing \eqref{eq:asymmetric} with $m=3,m'=2$.
	The order among the four bounds holds because $h_{1}(\lambda)\leq h_{2}(\lambda)\leq h_{3}(\lambda)$ and $\Theta$ is increasing in its first argument. 
 The comparative statics results for them can be obtained by differentiating each with respect to $\eta$ and then, again, applying that $h_{1}(\lambda)\leq h_{2}(\lambda)\leq h_{3}(\lambda)$ and $\Theta$ is increasing in its first argument.
\end{proof}


\cref{thm:asymmetric:group} implies that discrimination against the minority is more likely to take place when the total measure of buyers, $Q$, is large, as opposed to discrimination against the majority. It also suggests that discrimination is more likely when the two groups are more asymmetric in population size: As $\eta$ rises, the interval $(\underline{Q}_{1},\overline{Q}_{2})$---the set of dicrimination-compatible $Q$'s---expands. To obtain an intuition, consider the discriminatory equilibrium with the same group size and suppose $Q$ is slightly below $\overline{Q}$, where $\overline{Q}$ is defined in \Cref{thm:discriminatory_existence}. If $Q$ increases slightly above $\overline{Q}$, then the congestion effect dominates, so the discriminatory equilibrium unravels. Now, increase  the size of the advantaged group.  
This could weaken the congestion effect sufficiently so that the feedback loop  and the resulting discriminaroty equilibrium reemerge even with this larger $Q$. 
 Conversely, for $Q$ around $\underline{Q}$, where $\underline{Q}$ is defined in \Cref{thm:discriminatory_existence}, \emph{reducing} the size of the \emph{disadvantaged} group weakens the congestion effect, so discrimination can be sustained even below $\underline{Q}$, provided the group sizes are sufficiently asymmetric.

\newpage
\bibliography{CKZ_references1}

\end{document}